\documentclass[10pt,letterpaper]{article}
\usepackage[top=0.85in,left=2.75in,footskip=0.75in]{geometry}

\usepackage{amsmath,amssymb}

\usepackage{changepage}

\usepackage{textcomp,marvosym}

\usepackage{cite}

\usepackage{nameref,hyperref}

\usepackage[right]{lineno}

\usepackage[nopatch=eqnum]{microtype}
\DisableLigatures[f]{encoding = *, family = * }

\usepackage[table]{xcolor}

\usepackage{array}

\newcolumntype{+}{!{\vrule width 2pt}}

\newlength\savedwidth

\newcommand\thickhline{\noalign{\global\savedwidth\arrayrulewidth\global\arrayrulewidth 2pt}%
\hline
\noalign{\global\arrayrulewidth\savedwidth}}

\raggedright
\usepackage[aboveskip=1pt,labelfont=bf,labelsep=period,justification=raggedright,singlelinecheck=off]{caption}

\makeatletter
\renewcommand{\@biblabel}[1]{\quad#1.}
\makeatother

\usepackage{lastpage,fancyhdr,graphicx}
\usepackage{epstopdf}
\fancyheadoffset[L]{2.25in}
\fancyfootoffset[L]{2.25in}
\usepackage{soul}

\begin{document}
\vspace*{0.2in}

\begin{flushleft}
{\Large
\textbf\newline{Role of connectivity anisotropies in the dynamics of cultured neuronal networks} 
}
\newline
\\
Akke Mats Houben\textsuperscript{1,2,*}, Jordi Garcia-Ojalvo\textsuperscript{3}, Jordi Soriano\textsuperscript{1,2,*}
\\
\bigskip
\textbf{1} Departament de F\'{i}sica de la Mat\`{e}ria Condensada, Universitat de Barcelona, E-08028 Barcelona, Spain
\\
\textbf{2} Universitat de Barcelona Institute of Complex Systems (UBICS), E-08028 Barcelona, Spain
\\
\textbf{3} Department of Medicine and Life Sciences, Universitat Pompeu Fabra, E-08003, Barcelona, Spain
\\
\bigskip

%
%





* akkemats.houben@ub.edu, * jordi.soriano@ub.edu

\end{flushleft}
\section*{Abstract}

Laboratory-grown, engineered living neuronal networks {\em in vitro} have emerged in the last years as an experimental technique to understand the collective behavior of neuronal assemblies in relation to their underlying connectivity. 
An inherent obstacle in the design of such engineered systems is the difficulty to predict the dynamic repertoire of the emerging network and its dependence on experimental variables. 
To fill this gap, and inspired on recent experimental studies, here we present a numerical model that aims at, first, replicating the anisotropies in connectivity imprinted through engineering, to next realize the collective behavior of the neuronal network and make predictions. 
We use experimentally measured, biologically-realistic data combined with the Izhikevich model to quantify the dynamics of the neuronal network in relation to tunable structural and dynamical parameters. 
These parameters include the synaptic noise, strength of the imprinted anisotropies, and average axon lengths. 
The latter are involved in the simulation of the development of neurons {\em in vitro}. 
We show that the model captures the behavior of engineered neuronal cultures, in which a rich repertoire of activity patterns emerge but whose details are strongly dependent on connectivity details and noise. 
Results also show that the presence of connectivity anisotropies substantially improves the capacity of reconstructing structural connectivity from activity data, an aspect that is important in the quest for understanding the structure-to-function relationship in neuronal networks. 
Our work provides the {\em in silico} basis to assist experimentalists in the design of laboratory {\em in vitro} networks and anticipate their outcome, an aspect that is particularly important in the effort to conceive reliable brain-on-a-chip circuits and explore key aspects such as input-output relationships or information coding.

\section*{Author summary}

The computational modeling of living neuronal networks has become an important tool to help design experiments, test analytical methods and make predictions. 
Based on the growing interest in engineering neuronal circuits in vitro, here we provide a numerical model of neurons growing in anisotropic substrates, where neurons and connections are guided to grow in specific regions or follow pre-established paths. 
The model combines an algorithm to obtain biologically realistic connectivity, together with the Izhikevich neuronal dynamics to replicate the collective behavior of neuronal networks. 
The model reproduces well experimental observations and is able to make important predictions, such as the impact of development or synaptic noise in network dynamical traits. 
The model also demonstrates that the presence of anisotropies, and inherited constraints in neuronal connectivity, facilitates the reconstruction of structural connectivity from dynamics, a feature that is important to help designing in vitro systems aimed at approaching brain architectural and dynamical traits.


\section*{Introduction}
\emph{In vitro} cultured neurons are a widely used experimental model system to study the dynamics of complex networks of active excitable elements~\cite{millet2012over,keller2019past}. 
The accessibility and ease of manipulation of neurons in culture allow the {\em ad hoc} design of different network configurations and  physicochemical interventions, while monitoring single cell behavior in relation to emergent collective properties~\cite{aebersold2016,soriano2023}.
This malleability and observation potential enables to study questions such as plasticity~\cite{bakkum2008b, jimbo1999, wagenaar2006}, signal processing and propagation~\cite{duru2023, feinerman2005, feinerman2006}, the interchange between segregated and synchronized activity~\cite{yamamoto2018, yamamoto2023}, and the dynamical changes under pathological states~\cite{didomenico2019, wagenaar2005}, up to more abstract phenomena such as learning and memory~\cite{bakkum2008a, kagan2022, shahaf2001}.
The versatility and range of applicability of neuronal cultures have favored their extensive use to model universal phenomena in laboratory living neuronal circuits and the brain~\cite{aebersold2016, eckmann2007, soriano2023}.

However, cultured neuronal networks typically display a relatively poor dynamical behavior, such as dichotomous dynamics characterized by culture-wide coherent activity events ({\em network bursts}) interspersed with sparse random activations~\cite{kamioka1996, maeda1995}. 
This is very different from the rich activity repertoire of the living brain~\cite{deco2015}, and more closely resembles pathological states like epilepsy~\cite{wagenaar2005}. To enrich network dynamics {\em in vitro}, micro-engineered structures can be used to restrict and guide the connections between subpopulations of neurons, imprinting anisotropies such as modular organization and directed connectivity that promote the emergence of more diverse activity patterns~\cite{aebersold2016, bisio2014, pan2015, shein2011, yamamoto2018, yamamoto2022}. 

In a recent work by Montal\`{a}-Flaquer {et al.}~\cite{montalaflaquer2022}, polydimethylsiloxane (PDMS) topographical patterns were used to create an inhomogeneous growth environment that introduced subtle modulations in axon growth directions. The advantage of such mild modulations to realize modular networks in a bottom-up manner, as compared to strict confinement of neurons and connections, is that topographical modulations balance a coarse guidance of connections with network-wide self-organization, thus promoting the development of mesoscopic architecture without imposing rigid microscopic blueprints, a concept that was initially explored in Refs~\cite{okujeni2017mesoscale,okujeni2023structural}. 
In the study of Montal\`{a}-Flaquer et al.~\cite{montalaflaquer2022}, the topographical shapes came in two variants: (i) consistently interspaced parallel tracks and (ii) randomly placed squares. 
The introduction of these patterns infused the neuronal cultures with more diverse spontaneous activity repertoires, resembling those of neuronal circuits in the active brain.
A key feature of these subtly modulated patterns is that they facilitate local connectivity without fully suppressing network-wide activity, leaving the integration-segregation capacity of the neuronal networks~\cite{yamamoto2018} intact.

The above-described ability to modulate and guide the formation of connections in neuronal cultures opens up a vast array of experimental and engineering possibilities.
However, experiments take time to plan and execute, so it would be advantageous to be able to use an \emph{in silico} model to quickly explore possible PDMS structures or experimental manipulations.
In addition, having a numerical model of inhomogeneous neuronal cultures also allows to manipulate aspects which are difficult to control in biological neuronal cultures, and gives insight into aspects that are not directly observable experimentally, most prominently the structural connectivity of the neurons.
In that way, having access to a computational model aids to better understand phenomena observed in experiments.

Following this line of reasoning, in this paper we present a numerical model of neuronal cultures grown in topographically patterned environments. 
First we show that the model agrees with existing experimental observations~\cite{montalaflaquer2022}.
Next, the numerical simulations are used to (i) investigate the relationship between structural connectivity and dynamics, (ii) elucidate the effects of spatial anisotropy and noisy driving on the dynamics of the neuronal cultures, and (iii) to shed light on the ease of reconstructing the connectivity from recorded spontaneous activity under different conditions.

\section*{Results}

\begin{figure}
\center
\includegraphics[width=4.5in]{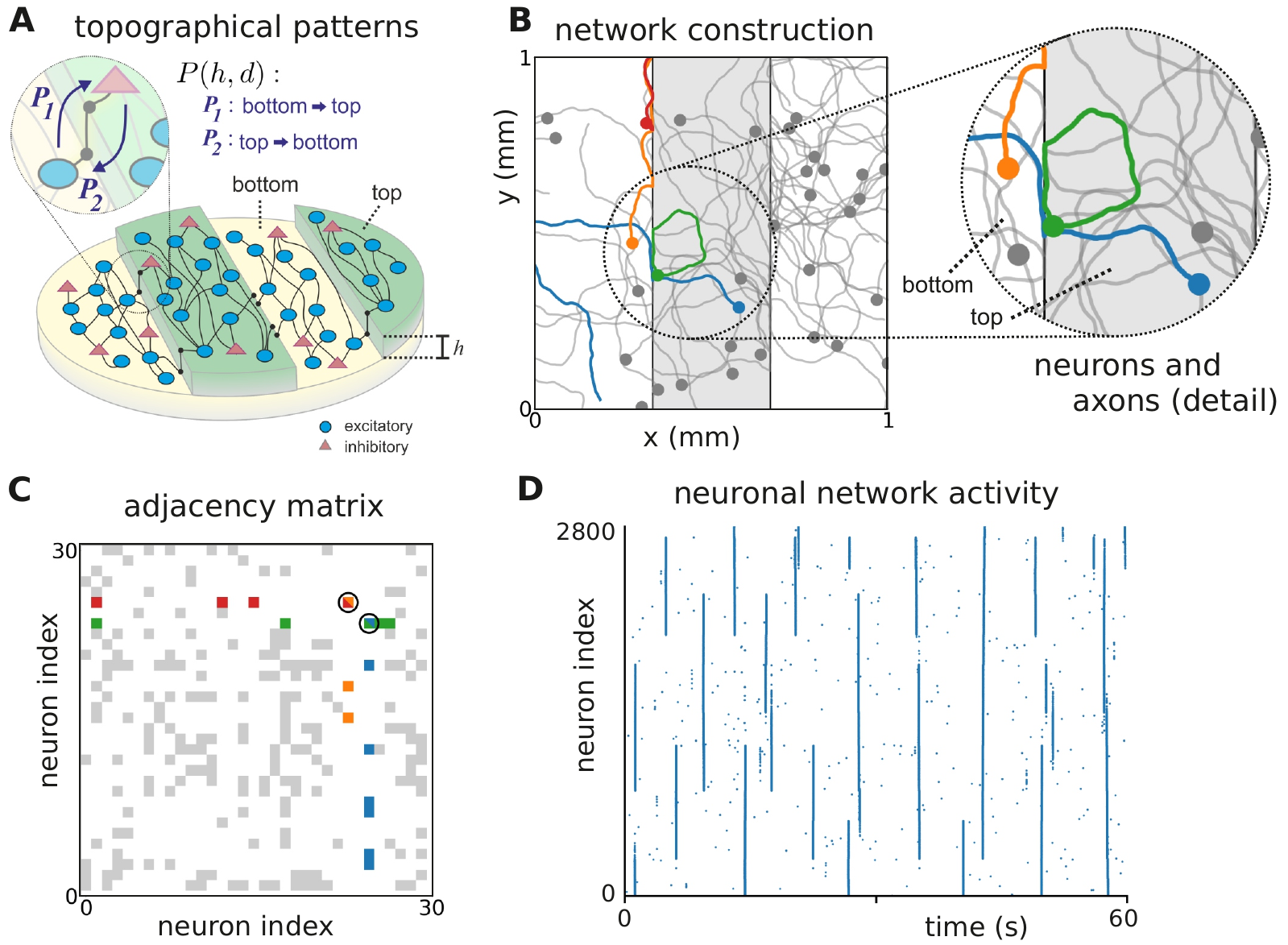}
\caption{{\bf {\em In silico} construction of neuronal networks with topography.} 
{\bf (A)} Conceptual representation of a topographical network, in which a mixed population of excitatory and inhibitory neurons connect following the tracks at the bottom or at the top of a profile of height $h$. The axons of neurons may pass into other tracks with probabilities $P_1$ and $P_2$. 
{\bf (B)} Illustrative network layout as implemented by the {\em in silico} model. Only 30 neurons are shown for clarity. Colored dots and lines illustrate the location of neuronal bodies and the excursions of their axons. The blue axon crosses from top to bottom, while the orange, red and green follow the edges of the pattern. 
{\bf (C)} Corresponding connectivity matrix, highlighting the interconnectivity between the colored neurons.
\textbf{(D)} Resulting network activity displayed as a rasterplot, in which each blue dot indicates the activation of a neuron (indexed along the vertical axis) along time (horizontal axis).
}
\label{fig:axon}
\end{figure}

We carried out an {\em in silico} exploration of the growth and activity of neuronal cultures in inhomogeneous environments.
The inhomogeneities consisted of topographical modulations of height $h$ in the substrate where neurons sit, illustrated in Fig~\ref{fig:axon}. 
As described in Methods, the {\em in silico} approach consists of two stages. 
First, the neuronal connections are established by simulating axon growth in two-dimensions whilst modeling the effect of PDMS obstacles on the direction of axon growth (Figs~\ref{fig:axon}A and B). This stage determines the network connectivity matrix (Fig~\ref{fig:axon}C). 
Secondly, spontaneous activity on the generated neuronal networks is simulated using the Izhikevich model (with 80\% excitatory neurons and 20\% inhibitory ones), and the obtained spike times for the entire network population (Fig~\ref{fig:axon}D) are stored for further analysis.

The parameters governing the neuronal dynamics were set such that the flat \emph{Control} condition, i.e., with $h=0$, reproduced qualitatively the dynamics observed in experiments on a flat surface~\cite{montalaflaquer2022}, characterized by strong network bursting events in which all neurons activated together in a short time window, or remained silent. 
The effect of $h>0$ on the crossing probabilities was fitted to experimental data~\cite{hernandeznavarrothesis} that quantified the likelihood that neuronal activity occurring at an area with $h=0$ could extend to another area with $h>0$, as described in the Methods. 
The crossing probabilities are given in Table~\ref{tbl:Pcross}, and the resulting parameter values for the neuronal dynamics are listed in Table~\ref{tbl:params}.

\subsection*{Dynamics of homogeneous and anisotropic {\em in silico} cultures}\label{sec:res_ncdyn}
\begin{figure}
\includegraphics[width=\textwidth]{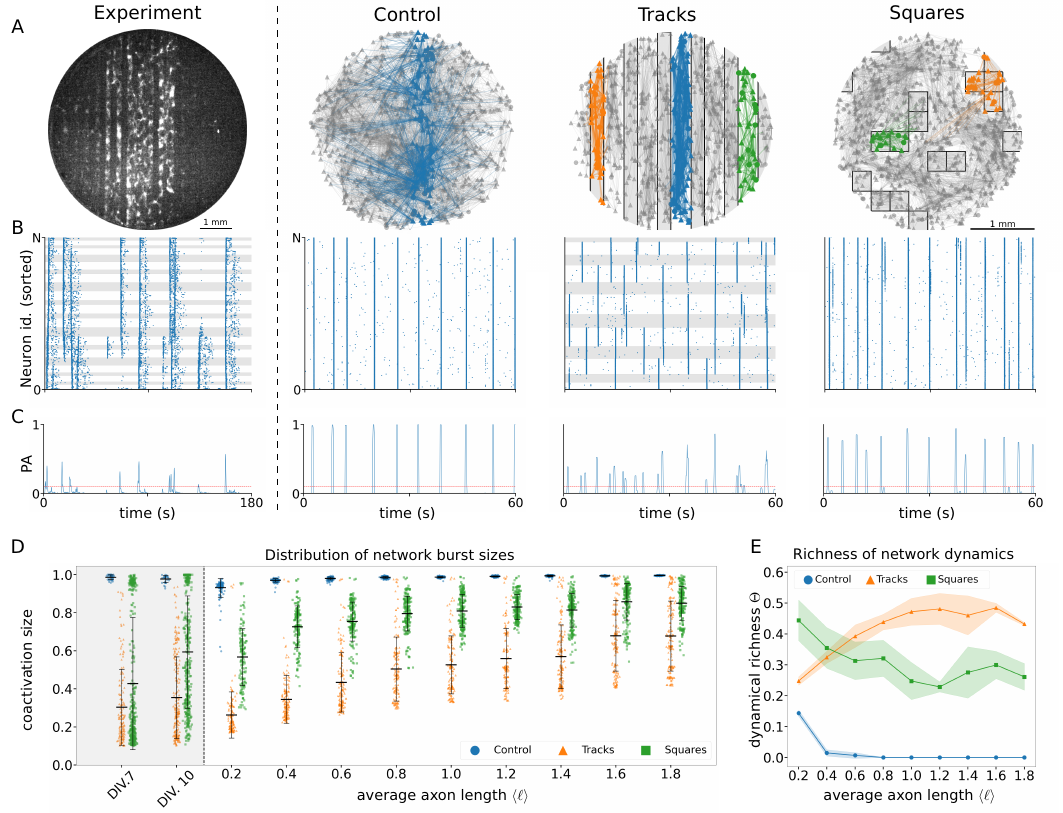}
\caption{{\bf Spontaneous activity in homogeneous and topographically-patterned networks.}
\textbf{(A)} Left panel shows a fluorescence image of an {\em in vitro} neuronal network, with bright objects corresponding to active neurons. 
The remaining panels show examples of simulated neuronal culture layouts and connectivity. 
Only 25\% of neurons and their connections are shown for clarity. 
For each simulated condition a group of neurons and their outgoing connections are highlighted in a different color in order to illustrate the effect of the PDMS obstacles.
\textbf{(B)} Raster plots of the networks in (A). 
Each dot represents a spike of a neuron. 
Neurons are sorted by their position along the left-right direction, with the leftmost neuron at the bottom ($0$) and the rightmost neuron on the top ($N$). 
Time is represented along the horizontal axis.
In the first and third columns, corresponding to the \emph{Tracks} condition in experimental and simulated data respectively, the background shading separates the neurons into the different tracks.
\textbf{(C)} Population activity ($PA$) plots showing the summed activity traces corresponding to the raster plots of panel B. 
The value `$1$' corresponds to the full culture being active in a short time window.
Only $1~\text{minute}$ sections of full simulations are shown for clarity.
\textbf{(D)} Distribution of the sizes of network bursts along development for different conditions.
For each value of $\langle \ell \rangle$, indicated along the $x$-axis, the peaks of the summed activity traces (see panel C) are plotted as points.
The two grey-shaded plots on the left correspond to experimental results from~\cite{montalaflaquer2022}.
For each axon length the peaks are plotted for the three conditions next to each other: \emph{Control} (blue circles), \emph{Tracks} (orange triangles) and \emph{Squares} (green squares). 
The figure shows network burst sizes of three independent simulation runs, each with a length of $5~\text{minutes}$ of simulation time, for varying axon length $\langle \ell \rangle$ and for the three experimental setups shown in (A).
\textbf{(E)} Richness of network activity shown in (D) for different axon lengths and setups.
Each line corresponds to a single condition: \emph{Control} (blue), \emph{Tracks} (orange) and \emph{Squares} (green).
Each dot indicates the average over 3 independent simulations runs, each with a length of $5~\text{minutes}$.
}
\label{fig:fig2}
\end{figure}

In order to validate the presented model, the simulated activity is compared to previously published experimental results. Fig~\ref{fig:fig2} summarizes the network connectivity and resulting activity dynamics of the model in the three conditions studied experimentally in~\cite{montalaflaquer2022} (\emph{Control}, \emph{Tracks} and \emph{Squares}), rendering it possible to compare the numerical simulations to experimental data. 
The first column of Figs~\ref{fig:fig2}A-C provides an example of the characteristic results obtained in biological cultures grown on PDMS tracks, showing a snapshot of the dynamics of the network together with the raster plot and the population activity (PA), i.e., the fraction of active neurons within a small time window. 
In the plots one can observe that the dynamics encompasses different groups of highly coordinated neurons, from small assemblies to the entire network, in contrast with a standard culture grown on a flat surface in which all neurons activate together or remain silent otherwise.

In a flat growth environment (\emph{Control} condition; second column of Fig~\ref{fig:fig2}A-C), axon positioning is unrestricted by obstacles. 
As such, neurons are able to establish a dense patchwork of connections, and can have long range projections that may reach the full width of the culture. This leads to culture-wide synchronized activity interspersed with periods of low activity (Fig~\ref{fig:fig2}B and C), as is characteristic for standard, unperturbed neuronal cultures~\cite{kamioka1996, maeda1995, montalaflaquer2022}.

In contrast, in the \emph{Tracks} condition (third column of Fig~\ref{fig:fig2}A), the obstacles lead to an anisotropic connectivity profile with abundant long-range connections along the tracks and few short-range connections traversing them. From this we expect that a track can be seen as a densely connected neuronal module that is weakly coupled to its neighboring tracks.
Indeed, from the raster plot (Fig~\ref{fig:fig2}B, third column) we observe that the periods of synchronized activity now consist of groups of neurons with a large variability in sizes: either single tracks, several temporarily synchronized tracks, or the full culture, are contemporaneously active.
The network population activity (Fig~\ref{fig:fig2}C) highlights this diversity of co-activation sizes. The activity dynamics of the \emph{Tracks} condition are indeed fitting to that of weakly interdependent modules. 
Each module seems to follow a dynamics similar to that of the full \emph{Control} culture, albeit at a smaller scale. 
We note that the modules sporadically synchronize with each other, leading to larger network bursts and a resetting of the phases of the synchronized tracks.

Lastly, the \emph{Squares} condition emerges as an intermediate state between the \emph{Control} and \emph{Tracks} cultures in terms of structural connectivity. Indeed, a large area exists in which axon growth is unobstructed, leading to a large group of neurons that can interconnect as in the \emph{Control} condition.
However small patches of largely isolated neurons constitute separate modules, weakly connected to the large interconnected group and potentially to other modules. 
This connectivity, illustrated in the last column of Fig~\ref{fig:fig2}A, leads to activity dynamics that is also intermediate between the \emph{Control} and \emph{Tracks} conditions, as expected. This is clear by comparing, e.g., the second and third columns with the last column of Fig~\ref{fig:fig2}B and C.
In most of the cases, synchronized activity encompasses the whole culture, or a large fraction of it.
Occasionally, the neurons of one or several of the largely isolated modules collectively activate in isolation from the rest of the culture.

\subsubsection*{Network development}
In the experimental study it was observed that along development (number of days {\em in vitro} (DIV) since plating the neurons on the glass cover slips), the distribution of network burst sizes changed in a distinct manner for the different conditions~\cite{montalaflaquer2022}. 
From the first day that activity was detectable, the \emph{in vitro} \emph{Control} cultures tended to synchronize fully during each network burst, a highly rigid dynamics that was observed until the last recording day (blue dots in the grey-shaded plots of Fig~\ref{fig:fig2}D).
In contrast, the \emph{Tracks} and \emph{Squares} \emph{in vitro} cultures both exhibited a wider distribution of burst sizes, that changed along development. 
As shown in the grey-shaded plots in Fig~\ref{fig:fig2}D, the \emph{in vitro} \emph{Tracks} condition started off at DIV 7 with a distribution heavily biased towards small numbers of neurons co-activating, which broadened to also include larger groups of co-activating neurons at the more developed stage DIV 10.
The \emph{in vitro} \emph{Squares} condition plays an intermediate role, with a nearly bimodal distribution for young cultures at DIV 7 displaying small bursts alternating with large culture-wide bursts. 
For older \emph{in vitro} \emph{Squares} cultures at DIV 10, the small bursts disappeared, showing mainly intermediate to large co-activation sizes.

In light of these results, it was proposed~\cite{montalaflaquer2022} that one main aspect governing the development of neuronal cultures after plating is the growth of axons, leading to a gradual increase in the amount and range of neuronal projections and, therefore, a higher probability for a neuron to connect with any other in the network.
Thus, as a proxy to model network development in the numerical simulations presented here, simulation runs were carried out for progressively higher values of the parameter $\langle \ell \rangle$ that determines the average axon length during the network growth phase.
The distributions of network burst sizes for the three network conditions and different values of $\langle \ell \rangle$ are shown in Fig~\ref{fig:fig2}D.
For the \emph{Control} and \emph{Tracks} conditions, the distributions show a development very similar to the experimentally observed distributions, with the \emph{Control} condition already displaying culture-wide synchronization for very short average axon lengths $\langle \ell \rangle = 0.2~\text{mm}$, and the \emph{Tracks} condition starting off concentrated at small co-activation sizes with a widening of the distribution for increasing $\langle \ell \rangle$.
The \emph{Squares} condition shows intermediate distributions between the \emph{Control} and \emph{Tracks} conditions, although the bimodal character of the distribution seen in experiments is not observed in the numerical simulations. This is possibly due to the fact that, in the experiments, a small group of isolated neurons may strengthen their connections to spontaneously activate, a phenomenon mediated by complex plasticity mechanisms that were not included in the simulations.

An alternative representation of the distribution of coactivation sizes consists in summarizing the burst sizes for each simulation run into a single number that captures the `dynamical richness' $\Theta$ of the activity. This measure was proposed and used to quantify the difference in activity patterns for different neuronal culture setups~\cite{yamamoto2018}. The computation of $\Theta$ from the raster plot data is described in the Methods section.
For the simulated data, the dynamical richness, shown in Fig~\ref{fig:fig2}E, displays interesting dependencies on the average axon length $\langle \ell \rangle$, distinct for each condition.
The \emph{Control} and \emph{Squares} conditions exhibit their dynamically richer activity for short axon lengths $\langle\ell\rangle = 0.2~\text{mm}$, with the value $\Theta$ decreasing monotonically for increasing $\langle \ell \rangle$.
In the \emph{Control} condition $\Theta$ vanishes for short average axon lengths $\langle \ell \rangle \approx 0.8~\text{mm}$, since each coactivation encompasses the whole network already for short axon lengths.
The \emph{Squares} condition has an overall higher dynamical richness because of the presence of isolated modules, which cause variation in the number of neurons that coactivate in each network burst.
The \emph{Tracks} condition displays a non-monotonic relation between dynamical richness $\Theta$ and $\langle \ell \rangle$, with optimal dynamically rich activity for intermediate average axon lengths $\langle\ell\rangle$ and decreasing $\Theta$ on both extremes.
This behavior can be interpreted as resulting from the large connectivity anisotropy of the system. 
This anisotropy leads to rich dynamics when the average axon length is large enough to excite multiple neurons within the same track, but not too large as to consistently induce collective activity in multiple tracks simultaneously. 

\subsubsection*{Initiation and propagation of fronts}
\begin{figure}
\includegraphics[width=\textwidth]{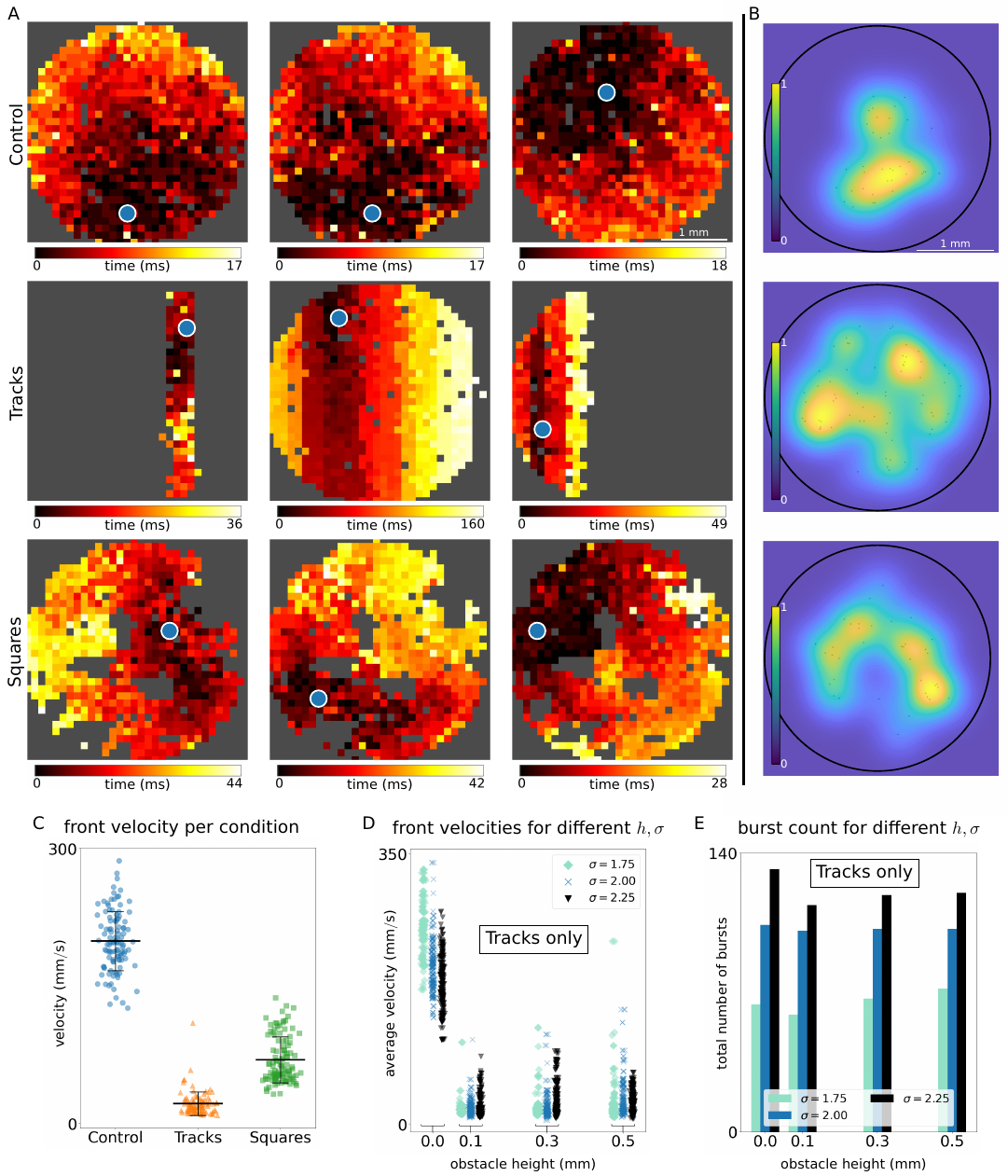}
\caption{{\bf Front propagation in different conditions.} 
    \textbf{(A)} Representative activity propagation through the neuronal cultures under different conditions. 
    Color-coding indicates first activation time of neurons in the spatial grid point within the front.
    Start of the activity wave is indicated with a white dot.
    \textbf{(B)} Spatial distribution of network burst initiation points (black dots) and its probability density function (pdf, blue-yellow colormap) for the different conditions as indicated on the left of panel (A).
    \textbf{(C)} Activity propagation velocities of each network burst for the different conditions.
    \textbf{(D)} Network burst propagation velocities for different parameter values of the obstacle height $h$ and noise intensity $\sigma$ for the \emph{Tracks} conditions.
    \textbf{(E)} Barplot indicating total number of network bursts for the same parameter values as in panel (D). }
\label{fig:fig3}
\end{figure}

We have seen so far that the activity of neuronal cultures is characterized by network bursts. 
From a spatiotemporal point of view, a network burst consists of a propagating activity front, successively exciting neurons such that each neuron fires multiple spikes in a short time span, followed by a phase-waveback, in which each neuron falls silent due to the depletion of synaptic resources.
As such, it is illuminating to investigate the spatial propagation of activity fronts by considering the first activation time of each neuron within the burst (Fig~\ref{fig:fig3}A), the burst's initiation points (Fig~\ref{fig:fig3}B), and propagation velocities  (Fig~\ref{fig:fig3}C).

Fig~\ref{fig:fig3}A shows representative examples of propagating bursts for the \emph{Control}, \emph{Tracks} and \emph{Squares} conditions. 
A visual inspection of the patterns reveal that the details of activity propagation, as well as the area covered in each burst, differs across designs. 
In the \emph{Control} condition (first row in Fig~\ref{fig:fig3}A) activity propagates as uninterrupted quasi-circular fronts that rapidly cross the whole culture, which agrees with the observation that each burst encompasses the full network. 
For the \emph{Tracks} condition (second row in Fig~\ref{fig:fig3}A), and as discussed before, the dynamics is reminiscent of a set of transiently synchronizing weakly coupled modules, each module behaving as a stochastic oscillator. 
As such it is expected that for some network bursts, activity only propagates across a single track (first plot in row) or a low number (third plot in row) of them, and at other times that the activity propagates throughout the whole culture  (second plot in row).
Moreover, when the burst traverses several tracks, the activity propagates first on a short time-scale within each track (as shown by each track having almost the same color in the second and third plots of the row), passing subsequently on to the next track on a slower time-scale (indicated by the fact that each track has a different color in the plot).
Due to this, the total duration of each network burst is dependent on the spatial scale of the activity propagation. 
Activity propagating within one track or a low number of tracks takes much less time (first and thirds plots in row), than a front traversing the whole culture (second plot in row), as seen by different ranges of the colorbars in the three plots.
Lastly, the activity propagation in the \emph{Squares} condition (third row in Fig~\ref{fig:fig3}A) is similar to that of the \emph{Control} condition, yet the propagation reveals intricate paths around the PDMS obstacles. This intricacy emerges from the fact that propagation can occur within each square module, on the main surface, or switching from one to another, leading to long, serpentine-like excursions across the culture.  


In addition to front propagation, one can also study the distribution of burst initiation points, i.e., the Euclidean position in the culture where each burst originated. This analysis is shown in Fig~\ref{fig:fig3}B where, for clarity of representation, the distribution of initiation points is shown as a blue-yellow heatmap that portrays the initiation probability. In accordance with previous experimental results~\cite{orlandi2013,montalaflaquer2022}, we observe that for the \emph{Control} condition (top plot of Fig~\ref{fig:fig3}B) the initiation points are focused in a relatively small area, indicating strong similarity between all the network bursts.
More varied burst initiation sites are observed in the \emph{Tracks} (central plot) and the \emph{Squares} (bottom) conditions. The former displays the broadest spatial distribution of initiation points, illustrating the rich repertoire of activity patterns in the \emph{Tracks} condition, while the \emph{Squares} condition falls in between the \emph{Control} and \emph{Tracks} conditions.

Lastly, we consider the summary of front propagation velocities in the different conditions. Fig~\ref{fig:fig3}C shows that uninterrupted fronts (\emph{Control} condition) propagate, on average, at $v_{C} = 199 \pm 32~\text{mm}/\text{s}$, an order of magnitude faster than the obstructed fronts, with $v_{T} = 21 \pm 12~\text{mm}/\text{s}$ and $v_{S} = 69 \pm 25~\text{mm}/\text{s}$ for \emph{Tracks} and \emph{Squares}, respectively. 
The fronts in the \emph{Squares} conditions propagate faster than in the \emph{Tracks} one, highlighting again their intermediate role, albeit more similar to the \emph{Tracks} condition. 
We note that, for \emph{Tracks}, the measured velocities correspond to fronts propagating perpendicular to their orientation. The velocity within a single track is very fast and similar to the one observed in the \emph{Control} case.

\subsubsection*{Impact of obstacle height $h$ and internal noise intensity $\sigma$ on activity propagation}

Here we considered only the data with the \emph{Tracks} configuration, given the strong asymmetry the obstacles imposed in the directions parallel or perpendicular to tracks. For this data, and following the above analysis, the front velocity is effectually measured along the tracks' transverse direction only. 
 
In general, we observed an interesting dependence of the front propagation velocity on obstacle height $h$ and internal noise intensity $\sigma$. 
As shown in Fig~\ref{fig:fig3}D, in the absence of obstacles ($h=0~\text{mm}$) the \emph{Tracks} condition reduces to the \emph{Control} one, and therefore the fronts propagate with high velocities, for the three noise amplitudes. 
However, a slight increase to $h=0.1~\text{mm}$ results in an order of magnitude drop in front velocity for all noise amplitudes, which remains for $h>0.1~\text{mm}$.

Focusing on the effect of different noise amplitudes $\sigma$, Fig~\ref{fig:fig3}D shows that for $h=0~\text{mm}$ a higher noise intensity leads to lower propagation velocities. 
This can be understood as wavefront break-ups, due to some of the neurons just ahead of the wavefront being in a refractory state due to spontaneous activation. In contrast, for increased obstacle height $h\geq 0.1~\text{mm}$ we see that stronger noise leads to some fronts with increased velocities.
We hypothesise that this can be explained using the concept of a firing ``quorum''~\cite{hernandez2021}: the idea that a neuron needs a minimal number of contemporaneously incoming excitations in order to be activated.
Increased noise driving leads to an increase in neurons spontaneously activating.
Given the low amount of connections crossing from one track to the next such an increase in spontaneous activation is beneficial, by leading to either a higher number of active neurons in the originating track (in order to utilize all the traversing connections), or to the activation of some of the neurons in the receiving track.

Lastly, we observe that increasing the noise driving $\sigma$ leads to an increase in the total number of bursts, but that changing obstacle height $h$ does not strongly affects the number of bursts (Fig.~\ref{fig:fig3}E).

\subsection*{Structural connectivity traits that shape network dynamics}
It is clear from the previous sections that different growth conditions affect the activity dynamics of the resulting culture differently, and they do so through their influence on the network connectivity.
A major advantage of numerical simulations over experiments is that one can access the structural network connectivity directly, a feature normally not accessible in experiments.
In this section we look at the structural connectivity matrices resulting from the network growth algorithm for $h=0.1~\text{mm}$ and graph-theoretical measures derived from these matrices.

\begin{figure}
\includegraphics[width=\textwidth]{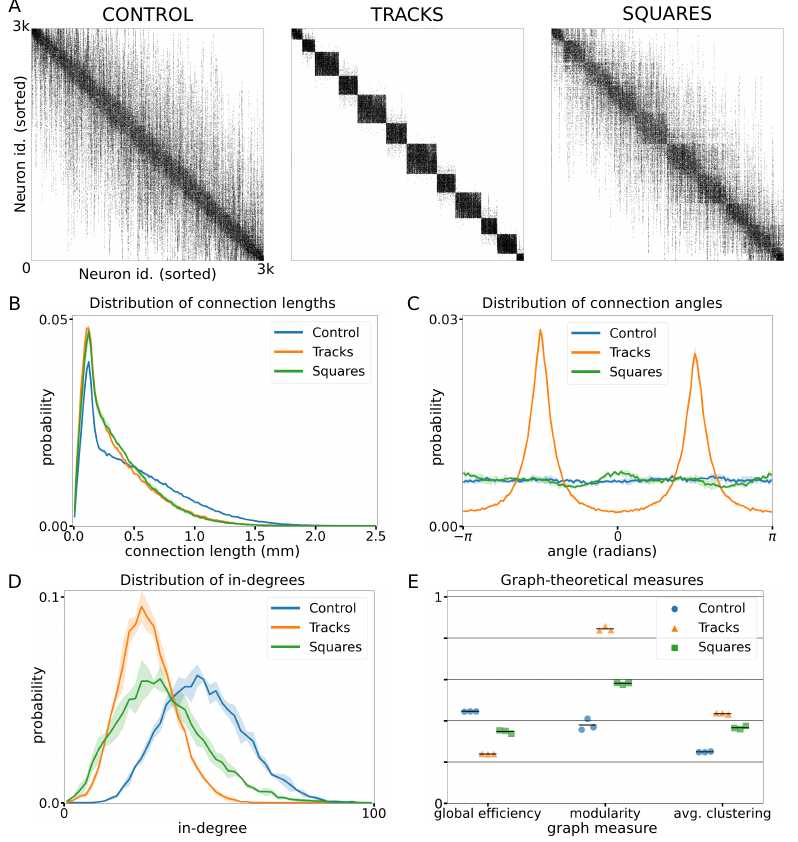}
\caption{{\bf Structural connectivity of simulated neuronal culture growth.}
    \textbf{(A)} Representative structural connectivity matrices for the three conditions. 
    Black dots indicate that a connection exists between the neurons represented in the two axes.
    Neurons are ordered according to their position in the neuronal culture from left to right.
    \textbf{(B)} Distribution of connection lengths between neurons for the conditions.
    \textbf{(C)} Distribution of angles between connected neurons.
    \textbf{(D)} Distribution of number of incoming connections.
    \textbf{(E)} Several graph-theoretical measures (from left to right: global efficiency, modularity $Q$, and average clustering).
}
\label{fig:fig4}
\end{figure}

Illustrative structural adjacency matrices, with neurons sorted by their horizontal position, are shown in Fig~\ref{fig:fig4}A for each of the three conditions.
As expected, the \emph{Control} condition shows a distance-dependent connectivity profile, in which connections are most probable close to the diagonal, i.e. between nearby neurons, and are less probable off-diagonal, mirroring the distance dependent nature of the connection algorithm.
In contrast, the \emph{Tracks} condition shows a clear modular connectivity blueprint, with differently sized boxes along the diagonal indicating strongly connected modules of nearby neurons (within the same track), with few connections between neurons from neighboring tracks and nearly no connections extending beyond neighboring tracks.
Lastly, the \emph{Squares} condition displays again an intermediate state. Some modular structure is discernible as in the \emph{Tracks} condition, yet largely the unmodulated distance-dependent connectivity profile of the \emph{Control} condition is followed.
These results show that what we can observe in the dynamics of the network, such as module-level, integrated versus segregated dynamics, can be linked clearly to the characteristics of the structural connectivity matrices.

Besides inspecting the structural connectivity matrices qualitatively, it is useful to look at the differences between the three conditions using coarse-grained observables. 
Figures~\ref{fig:fig4}B-E show a number of network measures calculated from the structural connectivity matrices.
Looking at the distribution of connection lengths (Fig~\ref{fig:fig4}B) it is apparent that there are predominantly short-range connections, of typically $0.20~\text{mm}$, in all three conditions.
The \emph{Tracks} and \emph{Squares} conditions exhibit very similar distributions, with a short tail vanishing at about $1.5~\text{mm}$, showing that the presence of obstacles confines most of the connections to a relatively small space.
In contrast, the \emph{Control} condition shows a heavier and longer tail vanishing at $2~\text{mm}$.
The differences in the distribution of connection lengths conform to those of the neuronal dynamics in the different conditions, especially \textemdash and as expected\textemdash\ to the co-activation sizes (Fig~\ref{fig:fig2}D) and propagation velocities (Fig~\ref{fig:fig3}C).

Figure~\ref{fig:fig4}C shows the distribution of the angle between the projecting and receiving neuron for all connections for the different conditions. 
The \emph{Tracks} condition shows that there is a clear predominance of connections at $\pm \pi/2~\text{radians}$, which correspond to connections that remain confined to each track, whereas for the \emph{Control} and \emph{Squares} condition no such bias exists.

Given the differences in the distance that the axons traverse in the different conditions, as apparent from Figs~\ref{fig:fig4}A-B, it is expected that different conditions also lead to different in-degree distributions, since axons confined to smaller spaces pass less unique neurons to connect to.
Figure~\ref{fig:fig4}D shows the in-degree distributions for the three conditions.
In accordance with the expectation, the \emph{Tracks} condition shows a narrow distribution centered at a low number of connections. 
Both the \emph{Control} and \emph{Squares} conditions show distributions with similar shapes and  broader than the \emph{Tracks} condition, indicating a wider in-degree range, largely due to the presence of more high-degree nodes than in the \emph{Tracks} condition.
Moreover, the distribution of the \emph{Control} condition is shifted towards higher in-degrees, owing to the free growth of the axons compared to the slight confinement in the \emph{Squares} condition.

Lastly, Fig~\ref{fig:fig4}E shows three graph-theoretical measures calculated on the structural connectivity matrices. 
The first measure is the unweighted topological global efficiency, which is the inverse of the average path length between pairs of neurons. 
As expected, the global efficiency of the \emph{Control} case is the largest, whereas the \emph{Tracks} condition has the lowest global efficiency and the \emph{Squares} is again intermediate. 
However, differences between the three conditions are small, which can be understood by the fact that all three have a strong bias towards short-range connections (Figs~\ref{fig:fig4}A-B), and thus in all three conditions a neuron needs several steps on average to reach distant neurons.
The second measure shown in this plot is the modularity $Q$, which quantifies the degree of modular structure in the network.
The \emph{Tracks} condition displays the most modular structure, characterized by the highest modularity $Q$ value, followed by the \emph{Squares} and lastly by the \emph{Control} condition.
Finally, the average clustering coefficient measures the number of closed triangles versus the total number of triangles in the network.
More strongly clustered or modular networks are expected to have a higher average clustering coefficient, and as expected the \emph{Tracks} condition has the highest value. 
However, both the \emph{Squares} and \emph{Control} conditions have very similar values, indicating again the predominance of local connectivity.

\subsection*{Spatial anisotropy and external noise driving}
The activity dynamics shown above for the \emph{Tracks} condition shows that there are two (experimentally modifiable) parameters that drive the differences in the behavior of neuronal cultures: (i) the intensity $\sigma$ of the noise driving, which influences the ease with which neurons activate, and (ii) the obstacle height $h$, which determines the strength of the anisotropy. 
The latter leads to an increase in modularity of the network that can be interpreted as a source of a spatially quenched disturbance.
In this section, we first investigate more extensively the effect of these two parameters on the dynamics of the neuronal culture. 
In the following section we use those results to find a relationship between the dynamical regime of the neuronal culture and the accuracy of structural connectivity reconstruction from the neuronal activity, using generalised transfer entropy.

\subsubsection*{Dynamical richness under different levels of anisotropy and noise}
\begin{figure}
\includegraphics[width=\textwidth]{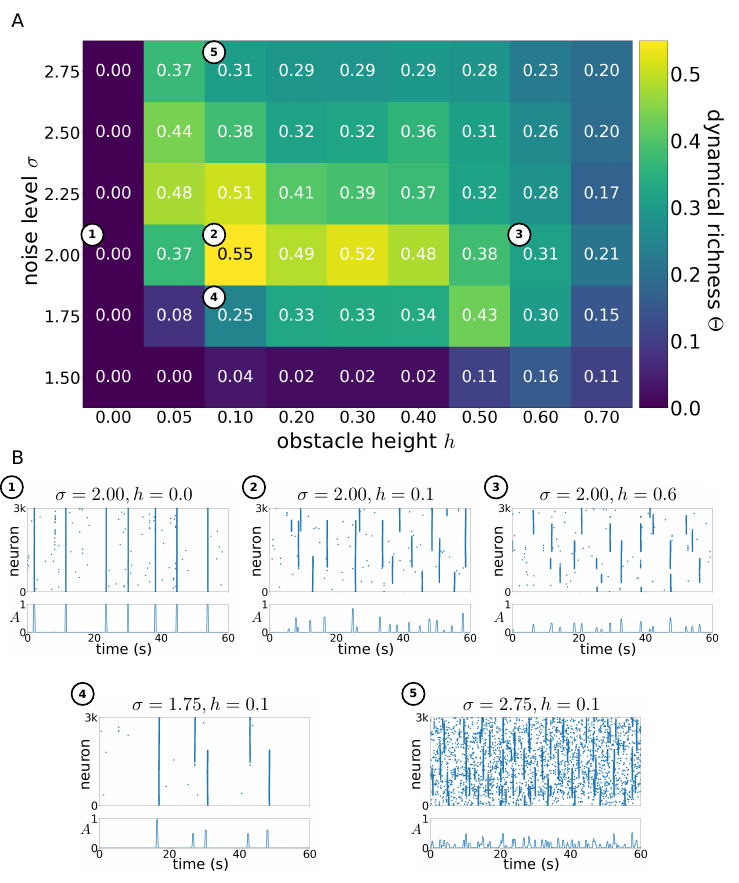}
\caption{{\bf Effects of spatial anisotropy and external noise intensity on dynamics.}
    \textbf{(A)} Dynamical richness $\Theta$ as a function of obstacle height $h$ and noise intensity $\sigma$ for the \emph{Tracks} condition.
    Data shown is averaged over three reproductions.
    White-circled numbers relate to highlighted plots in panel (B).
    \textbf{(B)} Representative raster-plots of the simulation runs with parameters as marked in panel (A), illustrating the differences in activity dynamics for the different parameter values and measured dynamical richness.
}
\label{fig:fig5}
\end{figure}

In order to understand the combined effect that the internal noise and the obstacle-driven anisotropy have on the dynamics, we calculate the dynamical richness $\Theta$ (see Methods) for different realizations of the neuronal culture as a function of the parameters $\sigma$ and $h$.
Figure~\ref{fig:fig5}A shows that intermediate values of both the noise intensity and the obstacle height $h$ (such as situation 2 in Fig~\ref{fig:fig5}) lead to a dynamically richer activity with a maximal repertoire of co-activation sizes. In contrast, for extreme parameter values the activity is either predominantly synchronized (situations 1 and 4 in the figure), or dominated by small-scale activations (situations 3 and 5).

The noise intensity $\sigma$ and obstacle height $h$ lead to differences in the co-activation dynamics on different scales, mirroring the differences in scale that these parameters affect. 
Comparing the raster plots of situations 1 and 3, which correspond to no spatial anisotropy ($h=0~\text{mm}$) and a very strong anisotropy ($h=0.6~\text{mm}$), respectively, we see that the synchrony-breakup induced by the obstacles occurs on the scale of single tracks.
In contrast, by comparing situations 4 and 5, corresponding to low ($\sigma=1.75$) and high ($\sigma=2.75$) noise intensities for the same obstacle height ($h=0.1~\text{mm}$), we observe that increased noise leads to the modules breaking up into randomly activating single neurons, with occasional synchronization of single modules.

Together, these results show that for both the external noise $\sigma$ and the spatial anisotropy $h$, optimal intensities exist that maximise the dynamical repertoire of the neuronal culture. These observations are important, since they may inspire experimentalists to design more elaborate {\em in vitro} networks that mimic brain-like dynamics, i.e. activity that is neither fully synchronous nor random. 
Additionally, the ability to bring a neuronal network to a state with an activity that is intrinsically rich appears as a key ingredient for optimal processing of input stimuli. 



\subsection*{Reconstruction of structural connectivity under different noise conditions }
The inference of structural connectivity between neurons from their recorded activity is still an open challenge, yet it is important for understanding the functioning of neuronal networks~\cite{Magrans2018, Banerjee2023}.
Transfer Entropy~\cite{schreiber2000}, and its extension for neuronal cultures called Generalised Transfer Entropy (GTE)~\cite{stetter2012, orlandi2014}, are extensively used tools to estimate the connectivity between neurons.
However, synchronized activity can act as a confounding factor in the estimation of neuronal connection strengths using TE~\cite{stetter2012, orlandi2014, magransdeabril2018}. 
For instance, when the network behavior is strongly dominated by culture-wide synchronous activity, TE over-estimates the number of connections.
In fact, GTE was proposed to deal with this by focusing the analysis on periods of non-synchronized activity~\cite{stetter2012}.
However, when strongly synchronized activity dominates the culture, there might not be sufficient data available for a reliable analysis.
In the previous section, it was shown that increased noise driving and spatial anisotropy lead to a desynchronization of the neuronal activations. Hence, a natural question arises whether GTE network inference works better for more modular $h>0$ and noisier $\sigma > 2.00$ {\em in silico} cultures.

\begin{figure}
\includegraphics[width=\textwidth]{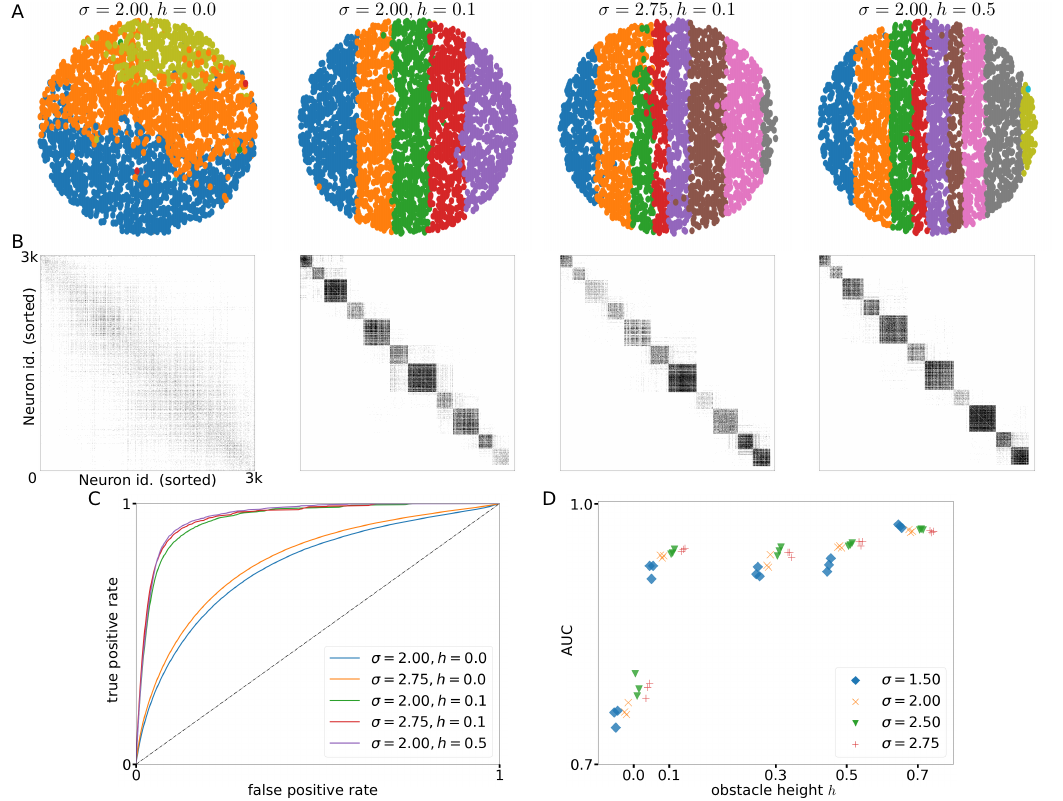}
\caption{{\bf Effective connectivity analysis.} 
    \textbf{(A)} Representative network maps, color-coded to indicate the functional modules that the neurons belong to, as found by estimating the effective connections of neurons using GTE and the Louvain algorithm, for different parameter values (different columns) for the \emph{Tracks} condition.
    \textbf{(B)} Effective connectivity matrices as given by thresholded GTE measures, corresponding to the networks in panel (A).
    Neurons are ordered as Fig~\ref{fig:fig4}A with the leftmost neuron at index $0$ and the rightmost neuron at $N$.
    \textbf{(C)} Receiver Operating Characteristic (ROC) curves for quantifying the resemblance between structural and effective network connectivity, for different parameter values.
    \textbf{(D)} Area Under the Curve (AUC) values of the ROC curves for a broad range of obstacle heights $h$ and noise amplitudes $\sigma$.
}
\label{fig:fig6}
\end{figure}

To address this question, we calculated the GTE values for the simulated neuronal cultures with different noise amplitudes $\sigma$ and obstacle heights $h$, keeping the condition $h=0$ as reference corresponding to a control network with no spatial anisotropies. 
Both the spatial network maps (Fig~\ref{fig:fig6}A), in which neurons are grouped into functional modules using the Louvain algorithm~\cite{blondel2008fast}, and the  effective  connectivity matrices (Fig~\ref{fig:fig6}B) show qualitative agreement with the structural network plots in Fig~\ref{fig:fig2}A and the structural connectivity matrices in Fig~\ref{fig:fig4}A.

In order to quantify the accuracy of the reconstruction of individual connections, we constructed Receiver Operating Characteristic (ROC) curves for the different parameter combinations.
As Fig~\ref{fig:fig6}C shows, the reconstruction quality separates into two groups.
For spatially isotropic cultures $h=0~\text{mm}$ (blue and orange lines in Fig~\ref{fig:fig6}C) the GTE method has a fair performance, with an area under the curve (AUC) near $0.8$ for all noise amplitudes, as shown in Fig~\ref{fig:fig6}D.
The presence of even a slight spatial anisotropy $h=0.1~\text{mm}$ increases the quality of the GTE reconstruction to an AUC value of $0.9$, with a very small increase in performance for higher noise amplitudes $\sigma$ (purple vs green line in Fig~\ref{fig:fig6}C).
Further increase of the obstacle height only slightly increases the reconstruction performance (red vs green line in Fig~\ref{fig:fig6}C).
Hence, primarily, the increased separation of the activity dynamics into modular tracks leads to an increased capacity to reconstruct the structural connectivity using GTE, more than promoting sparse activation through increased noise.
This is possibly explained by the fact that effective connections between pairs of neurons can exist mediated by several intermediate neurons, leading to very long effective connections between neurons that are not connected directly. 
The confinement of activity to largely separated tracks limits the spatial extent of effective connections to the same track, while at the same time the higher modularity of the structural connections increases the probability that any two neurons within the same track are connected.
In contrast, increasing the noise intensity decouples the activity of single neurons, which is an effect that can be both advantageous and detrimental to the reconstruction of structural connections from effective connections, since for noise to be advantageous the activity of neurons need to be decoupled from the global activity, yet smaller scale synchronization between connected neurons is needed for GTE to be able to measure their connection.
Fig~\ref{fig:fig6}C shows that for slight anisotropy $h=0.1~\text{mm}$ and a high noise intensity $\sigma=2.75$ the reconstruction is as good as for a very strong anisotropy $h=0.5~\text{mm}$ and a intermediate noise intensity $\sigma=2.00$ (compare red and purple lines in the figure), indicating that the presence of both spatial anisotropy and noise driving helps to decouple the activity of the neurons from culture wide synchrony.

The observation that spatial anisotropies are the primary ingredient in reconstruction accuracy is interesting. 
There has been abundant discussion in the literature on the capacity to precisely unravel the connectivity blueprint of a neuronal network from activity data only~\cite{menesse2024integrated,orlandi2014}, since the intrinsic nonlinear nature of neurons in combination with intrinsic noise and variability of neuronal types makes the \textit{inverse problem} of perfect reconstruction unattainable, requiring the incorporation of additional tools such as interrogation or labeling of neurons and connections. 
Our study shows that an even small restriction or guidance of neuronal connectivity by topographical anisotropy suffices to largely restrict the available connectivity repertoire and substantially improve reconstruction performance.


\section*{Conclusions}


In this work we constructed an {\em in silico} model to replicate the connectivity structure and dynamics of laboratory-grown neuronal cultures characterized by spatial anisotropies, which mould and dictate the layout of connections in the network~\cite{montalaflaquer2022}. 
The imprinted anisotropies, leading to non-uniform connectivity probabilities among neurons in spatially-confined regions, led to activity patterns that substantially departed from the control scenario of free connectivity. 
The simulated networks not only reproduced the experimental behavior, but also provided valuable insight on the role of the two key ingredients that govern collective behavior, namely the strength of the spatial anisotropy and the noise intensity that drives spontaneous activity.
The results show that, whereas the dynamics in control (non-anisotropic) networks is dominated by regular bursting episodes that encompass the entire system, the presence of inhomogeneities markedly favors a much richer repertoire of activity characterized by a broad range of co-activation patterns. 
However, both the spatial anisotropy strength and noise intensity have to be mild in order to maximize the richness of collective activity, otherwise the system is locked into extreme states of permanent bursting or random driving of activity in spatially isolated areas.
Thus, our work is inspirational for those studies, both {\em in silico} and {\em in vitro}, that aim at understanding, and even mimicking, the richness of brain-like dynamics and its relation to the network's building blocks, neurons and connections.  

In our study we considered both excitatory and inhibitory neurons, to mimic existing experimental designs~\cite{montalaflaquer2022}. 
The impact of the excitatory-inhibitory balance was not investigated in-depth in the present work. 
However, we carried out exploratory simulations in which inhibitory neurons were inactivated,  observing that network dynamics evolved towards stronger bursting. 
This result is in agreement with diverse experimental and numerical studies~\cite{orlandi2013, parodi2023deepening, yamamoto2023} describing that blocked inhibition reduces the capacity of the network to accommodate rich and diverse dynamical states, driving the network towards a pathological-like hyper-synchronous state. 
Since the balance between excitation and inhibition is altered in neurological diseases such as epilepsy and genetic forms of Parkinson's, which have been investigated {\em in vitro}~\cite{jablonski2021experimental, Carola2021}, our work may help experimentalists to engineer neuronal cultures that achieve a rich repertoire of activity patterns in healthy conditions to then explore the degradation in dynamics as inhibition is gradually lost. 

Finally, we observe that GTE-reconstructed connectivity approaches well the underpinned structural network in the case of anisotropic cultures. 
This evinces that imprinted spatial constraints favor the occurrence of overt neuron-to-neuron interactions which, as a counterbalance to network-wide bursting or random activity, is the key ingredient to bring to light the key topological and organizational traits of the structural connectivity. 
Our results fit well with the experimental work of Montal\`{a}-Flaquer {\em et al.}~\cite{montalaflaquer2022}, in which the {\em Tracks} configuration was seen to exhibit an abundance of effective connections oriented along the tracks themselves, a feature that was independently confirmed through immunostaining.
Thus, our study provides a strong {\em in silico} evidence to support the development of neuroengineered neuronal cultures, i.e., those systems where neurons and connections are tailored to fit {\em ad hoc} configurations and activity patterns. 
These configurations not only imprint rich dynamical traits that resemble brain dynamics, but their topological features can be predicted and analyzed. 
This may be crucial to understand the processing of input stimuli in a neuronal network, where a wealth of ingredients play an important yet elusive role, including network architecture, noise, and the balance between excitation and inhibition~\cite{haroush2019inhibition}.      




\section*{Materials and methods}
Numerical simulations aimed at replicating the two-level topographical patterns of the experiments described by Montal\`{a}-Flaquer {\em et al.}~\cite{montalaflaquer2022} in which, as sketched in Fig~\ref{fig:axon}A, a PDMS mold contained vertical modulations in the form of parallel tracks with a height $h$. 
Neurons (80\% excitatory and 20\% inhibitory) preferentially connected along the tracks either at the bottom or at the top of the mold. 
Neurons at the bottom could project connections to neurons at the top with some probability $P_{\text{bottom}\to\text{top}}$, and neurons from the top to the bottom with probability $P_{\text{top}\to\text{bottom}}$, effectually shaping a globally interlinked system yet with most connections along the tracks. 

To realize such a scenario {\em in silico}, neuronal networks were modeled in two-dimensional circular cultures with radius $r=1.5~\text{mm}$,
on which neurons were plated with a uniform density of $\rho = 400~\text{neurons} / \text{mm}^2$, resulting in $N=\lfloor \rho \pi r^2 \rfloor \approx 2800$ neurons in total (Fig~\ref{fig:axon}B).
The $N$ neurons of the network were placed randomly on a circular area in such a way that somas (circular areas of radius $r_{soma} = 7.5~\mu\text{m}$) of neighboring neurons did not overlap. 
The network connections were determined following the algorithm of Orlandi~\emph{et al.}~\cite{orlandi2013}, complemented with the influence of the PDMS topography on axon growth, as described below.

\subsection*{Growing the networks}\label{sec:axon}
Starting from the center of each neuron $i$, axon growth was modelled by concatenating line segments of length $\Delta \ell = 10~\mu\text{m}$ up to a total length $\ell_i$, which for each neuron was independently drawn from a Rayleigh distribution with mean $\mu_\ell$. 
Each consecutive segment was placed at an angle drawn from a Gaussian distribution with zero mean and standard deviation $\sigma_\phi = 0.1~\text{radians}$ with respect to the direction of the previous segment, as illustrated in Fig~\ref{fig:axon}B. 
The dendritic tree of each neuron $i$ was modelled as a circular area of radius $r^{(i)}_{\text{dr}}$, drawn from a Gaussian distribution with mean $\mu_{\text{dr}} = 150~\mu\text{m}$ and standard deviation $\sigma_{\text{dr}} = 20~\mu\text{m}$.

During the growth of the axons, whenever a line segment was placed such that it crossed the border of an obstacle, two things could happen depending on (i) the angle between the axon and the border, and (ii) the height of the obstacle.
On the one hand, if the angle between the axon line segment and the border was smaller than $30^{\circ}$, the segment was replaced by one parallel to the border plus a random deviation on the order of $\sigma_\phi$, as illustrated in  the detailed map of Fig~\ref{fig:axon}B (orange and green axons). 
On the other hand, if the angle between the line segment and the border was larger than $30^{\circ}$, the segment remained there with probability $P(h, d)$, to simulate the crossing from a top to bottom obstacle or vice versa (Fig~\ref{fig:axon}B, blue axon). 
The probability $P(h, d)$ depended on the height $h$ of the obstacle and the direction $d\in\{\text{bottom}\to\text{top},~\text{top}\to\text{bottom}\}$ that the axon followed when crossing the obstacle border. 
If this probability $P(h, d)$ was satisfied, the axon crossed the obstacle border and the line segment counted to have a length of $\Delta \ell + h~\mu\text{m}$. 
With complementary probability $1-P(h, d)$ the axon was again deflected, meaning that it was replaced by a segment parallel to the border. 
The numerical values used to quantify these probabilities of axonal crossing for different heights were obtained from experimental data~\cite{hernandeznavarrothesis} as explained below, and are provided in Table~\ref{tbl:Pcross}.

Once all axons were placed on the substrate, a connection between neurons $i$ and $j$ was established whenever the axon of neuron $i$ crossed the dendritic tree of neuron $j$ with a probability $\alpha=0.5$. 
The set of network connections was stored in the {\em structural} adjacency matrix ${\bf S}=\{ s_{ij}\}$, with $s_{ij}=1$ for the presence of a connection $j \to i$ and $s_{ij}=0$ otherwise (Fig~\ref{fig:axon}C).
The weighted connectivity matrix $\mathbf{W} = \{w_{ij}\}$ is obtained from the structural matrix $\mathbf{S}$ by drawing a connection weight from a uniform distribution for each existing connection, $s_{ij} = 1 \implies w_{ij} \sim U(0, 1)$.
The sign of the outgoing signal is determined by whether a neuron is excitatory (80\% of the network) or inhibitory (remaining 20\%), through the synaptic dynamics described in the next section.


\subsection*{Neuronal dynamics}
The dynamics of each neuron was governed by the Izhikevich two-dimensional quadratic integrate-and-fire model with adaptation~\cite{izhikevich2003}, which provides a good balance between computational efficiency and biological accuracy, and was previously used for numerically simulating neuronal cultures~\cite{alvarezlacalle2009, orlandi2013}. 
The model is given by
\begin{equation}
\begin{aligned} \label{eq:dv}
    \frac{dv_i}{dt} &= 0.04v_i^2 + 5v_i + 140 - u_i + \sum_{j=0}^N w_{ij} p_j + \sigma \eta_i, \\
    \frac{du_i}{dt} &= \epsilon(\rho v_i - u_i),
\end{aligned} 
\end{equation}
where $\mathbf{W} = \{ w_{ij} \}$ is the above-defined adjacency matrix of connections (directed and weighted) between neurons, and $\eta_i$ is a Gaussian white noise term that captures the sporadic activation of each neuron $i$. 
The parameter $\sigma$ quantifies the intensity of the white noise, and can be tuned to emulate the abundant spontaneous activity observed in biological neuronal networks. 

The membrane potential $v_i$ and recovery variable $u_i$ are reset once the membrane potential exceeds a threshold $v_c$ and the neuron is said to ``spike'':
\begin{eqnarray}
    v_i(t) > v_c \implies \begin{cases} 
        v_i & \leftarrow v_0\\
        u_i & \leftarrow u_i + \delta_u.
    \end{cases}
\end{eqnarray}
After these resets, the dynamics is again governed by Eq.~(\ref{eq:dv}).

Once a neuron elicits a spike, it sends signals to the neurons it is connected to. 
This signal is determined by the synaptic dynamics, governed by two variables, namely the synaptic potential $p_i$ and the synaptic recovery variable $q_i$. 
The synaptic potential $p_i$ follows the equations
\begin{eqnarray}\label{eq:dp}
    \tau_p^{(i)}\frac{dp_i}{dt} = -p_i, &&
    v_i(t) > v_c \implies p_i \leftarrow p_i + \delta_p^{(i)} q_i,
\end{eqnarray}
which describe the decay and release of neurotransmitters in the synaptic cleft, respectively, assuming the rising-phase of the neurotransmitter release to be instantaneous~\cite{destexhe1994}.
The decay time $\tau_p^{(i)}$ and reset constant $\delta_p^{(i)}$ depend on the type of neuron $i$ (excitatory or inhibitory), specifically the reset constant $\delta_p^{(i)}$ will be non-negative (non-positive) for excitatory (inhibitory) neurons.

The synaptic recovery variable $q_i$ accounts for the phenomenon of synaptic depression, according to which neurons that are tonically active experience a reduction in the efficacy of synaptic transmission~\cite{cohen2011, eckmann2007}. 
This phenomenon can be described by an evolution equation and reset rule given by
\begin{eqnarray}
    \tau_q \frac{dq_i}{dt} = 1-q_i, &&  
    v_i(t) > v_c \implies q_i \leftarrow (1-\delta_q) q_i.
\end{eqnarray}
The parameters used in the numerical simulations are listed in Table~\ref{tbl:params}.

\subsection*{Determination of the obstacle-crossing probabilities}\label{sec:fitpcross}
The probability of an axon crossing an obstacle wall is determined by using calcium imaging data of neuronal activity crossing a PDMS-glass border~\cite{hernandeznavarrothesis}.
The axon crossing probabilities are then fitted by comparing the numerically simulated activity to this data.

The data consists of recordings of different neuronal cultures, each $6~\text{mm}$ in diameter, prepared on glass but with half of the culture covered with a PDMS layer of height $h$. 
The set of prepared cultures had different heights between the glass and the PDMS.
This allowed to calculate the number of bursts that are initiated either in the glass part or the PDMS part of the plate, and to register the frequency at which the activity originating in either half propagates successfully across the glass-PDMS border.
The second and third column of Table~\ref{tbl:Pcross} show the crossing probabilities of the activity obtained for different PDMS heights.
\begin{table}
\begin{adjustwidth}{-2.25in}{0in} 
\centering
\caption{
{\bf Probabilities of activity and axons crossing a PDMS border.} 
} 
\begin{tabular}{|l||l|l||l|l|}
\hline
    & \multicolumn{2}{c||}{\bf Activity propagation} & \multicolumn{2}{c|}{\bf Axon crossing} \\\hline
    {\bf PDMS height $h~\text{(mm)}$} & {\bf bottom$\to$top} & {\bf top$\to$bottom} & {\bf $P_{\text{bottom}\to\text{top}}$ } & {\bf $P_{\text{top}\to\text{bottom}}$ } \\ \thickhline
    $0.0$   & $1$   & $1$   & $1$                   & $1$                   \\ \hline
    $0.1$   & $0.60$& $0.85$& $0.45\times 10^{-3}$  & $3.3\times 10^{-3}$   \\ \hline
    $0.4$   & $0.15$& $0.90$& $0.25 \times 10^{-3}$ & $3.3\times 10^{-3}$   \\ \hline
    $0.6$   & $0.05$& $0.30$& $0.02 \times 10^{-3}$ & $0.50\times 10^{-3}$  \\ \hline
    $>0.7$  & $0$   & $0$   & $0$                   & $0$                   \\ \hline
\end{tabular}
\label{tbl:Pcross}
\end{adjustwidth}
\end{table}

Subsequently, numerical simulations are run with different values for axons crossing from glass to PDMS $P_{\text{glass}\to\text{PDMS}}$ and PDMS to glass $P_{\text{PDMS}\to\text{glass}}$, and the activity propagation across the glass-PDMS border are registered and compared with the experimental values.
In case the fraction of activity crossing in a certain direction is too low (high) the corresponding axon crossing probability is increased (decreased) until the simulated activity propagation fraction agrees with the experimental ones within $5\%$ precision.
The obtained axon crossing probabilities are given in the last two columns of Table~\ref{tbl:Pcross}.

\subsection*{Measures of neuronal activity}
\subsubsection*{Population activity}
The population activity $PA(t)$ indicates the number of active neurons within a short time window $\tau_{\text{win}}$ centered around time $t$, divided by the total number of neurons:
\begin{equation*}
    PA(t) = \frac{1}{N}\left\lvert \left\{ i~:~\exists_{t^{(k)}_i} |t^{(k)}_i-t| < \frac{\tau_{\text{win}}}{2}  \right\} \right\rvert.
\end{equation*}
Here $t^{(k)}_i$ is the time of the $k$-th spike of neuron $i$ and $\tau_{\text{win}} = 200~\text{ms}$ the size of the rectangular window. 
Sharp peaks in $PA(t)$ reveal strong network co-activations, i.e. groups of neurons (from small ensembles to network bursts) that spike coherently within a short time window.

\subsubsection*{Dynamical richness}
Dynamical richness~\cite{zamora2016functional,yamamoto2018} $\Theta$ is a measure of the diversity of co-activation sizes.
In the present work, $\Theta$ is calculated by creating a $m$-bin histogram of the peaks $\Gamma_i$ of the population activity $PA(t)$ trace of a recording, resulting in an estimated distribution of peaks $p(\Gamma)$, and following calculating the deviation of $p(\Gamma)$ from a uniform distribution.
\begin{equation*}
    \Theta = 1- \frac{m}{2(m-1)}\sum_{i=0}^{m-1}\left\lvert p\left(\frac{i}{m} \leq \Gamma < \frac{i+1}{m}\right) - \frac{1}{m}\right\rvert.
\end{equation*}

The quantity $\Theta$ attains values in the range $[0,1]$.
Values of $\Theta$ close to $0$ indicate that the network operates at full extremes of either random neuronal activity or fully synchronous network bursts, while values of $\Theta$ close to $1$ indicate that all possible co-activation sizes are present in the system.


\subsection*{Network measures}

\subsubsection*{Effective connectivity}
Causal interactions between pairs of active neurons were computed through a GTE implementation~\cite{stetter2012} run in Matlab~\cite{MATLAB}. 
Specifically, pairs of neuronal activity trains $I$ and $J$ were constructed as binary series $600~\text{s}$ long with values of either `1' (presence of activity) or `0' (absence of activity), binned at $10~\text{ms}$. For computation, and following Ref.~\cite{stetter2012}, the Markov order was set to $2$ and instant feedback was present. 
An effective connection
between any pairs of neurons $I$ and $J$ in the network was then considered whenever the information contained in $I$
significantly increased the capacity to predict future states of $J$. 
For that, raw transfer entropy estimates $\text{TE}_{I \rightarrow J}$ were first obtained, and then compared with the joint distribution of all inputs $X$ to $J$ and all outputs $I$ to $Y$ (for any X and
Y), as
\begin{equation}
z_{I\rightarrow J} = \frac{\text{TE}_{I \rightarrow J}- \langle \text{TE}_{\textbf{joint}} \rangle}{\sigma_{\text{joint}}}. 
\end{equation}
Here $\langle \text{TE}_{\textbf{joint}} \rangle$ and $\sigma_{\text{joint}}$ are the average value of the joint distribution and its standard deviation, respectively.
A significance threshold $z_{\text{th}}=2$ was established to accept an effective interaction as significant, setting $z_{I\rightarrow J} = 1\; \forall\, z_{I\rightarrow J} \geq z_{\text{th}}$ and $0$ otherwise. 
A threshold $z_{\text{th}}=2$ was set to compute effective connectivity data in the same way as the reference experimental work of Montal\`{a}-Flaquer {\em et al.}~\cite{montalaflaquer2022}. 
The derived effective connectivity matrices $\textbf{E} = \{ e_{IJ}\}$ were therefore directed but binary, which allowed for a direct comparison with the underlying structural connectivity of the studied synthetic networks.

Throughout the text in the present work, the term `effective' is used to refer to TE-inferred connections among neurons and the derived connectivity matrices. 
The term `functional' is used to refer to the broader concept of  network organization and characteristics.

\subsubsection*{Modularity analysis}

We used the modularity statistic $Q$ to quantify the tendency of neurons to organize into groups (termed modules) that were strongly connected within themselves and sparsely connected with other groups~\cite{Newman2004Analysis}. 
This quantity is defined as: 
\begin{equation}
Q = \frac{1}{2m}\sum_{i,j} \left(e_{ij}-\frac{k_i k_j}{2m}\right)\delta(c_i,c_j),\, 
\label{eq:3}
\end{equation}
where $N$ is the number of neurons, $e_{ij}$ represents the effective connectivity matrix, $k_i = \sum_{j=1}^{N}T_{ij}$ is the sum of the connections attached to neuron $i$, $c_i$ is the community to which neuron $i$ belongs, $m = (1/2) \sum_{i,j=1}^{N} T_{ij}$, and $\delta (u, v)$ is the Kronecker Delta with $\delta (u, v) = 1$ for $u = v$ and $0$ otherwise. 
$Q$ varied between $0$ (the entire network is the only module) and $1$ (each neuron is a module), with intermediate values indicating the presence of modules of varying size. 
The optimal modular structure was computed using the Louvain algorithm~\cite{blondel2008fast}. 
Detected modules were color-coded in the network maps to investigate whether modularity was related to the structure of the spatial disorder.

\subsection*{Simulation parameters}
The parameters used for the simulations are listed in Table~\ref{tbl:params}. 
Simulations were run for typically $600~\text{s}$ to obtain sufficient co-activation events for statistics. 
For each condition ({\em Control}, {\em Tracks} or {\em Squares}), $3$ network realizations were considered.
\begin{table}
\begin{adjustwidth}{-2.25in}{0in} 
\centering
\caption{
{\bf Parameter values used for simulations.}}
\begin{tabular}{|c|c|c|l|}
\hline
{\bf Parameter} & {\bf value} & {\bf unit } & {\bf  }\\ \thickhline
    \multicolumn{4}{|l|}{Neuron dynamics} \\ \thickhline
    $\sigma$    &           & $\text{mV}^2\text{ms}$    & noise intensity               \\ \hline
    $\epsilon$  & $0.02$    & $\text{ms}$               & adaptation time-scale         \\ \hline
    $\rho$      & $0.2$     &                           & adaptation $v$-sensitivity    \\ \hline
    $v_c$       & $30$      & mV                        & voltage threshold             \\ \hline
    $v_0$       & $-65$     & mV                        & voltage reset value           \\ \hline
    $\delta_u$  & $6.5$     &                           & adaptation reset increment    \\ \thickhline
    \multicolumn{4}{|l|}{Synapse dynamics} \\ \thickhline
    $\tau_p^{E}$    & $10$  & ms        & excitatory synaptic time-scale            \\ \hline
    $\delta_p^{E}$  & $3$   & mV        & max. excitatory post-synaptic potential   \\ \hline
    $\tau_p^{I}$    & $10$  & ms        & inhibitory synaptic time-scale            \\ \hline
    $\delta_p^{I}$  & $-6$  & mV        & max. inhibitory post-synaptic potential   \\ \hline
    $\tau_q$        & $1 \times 10^3$   & ms     & synaptic depression time-scale            \\ \hline
    $\delta_q$      & $0.8$ &           & decrease of synaptic vesicle pool         \\ \hline
\end{tabular}
\label{tbl:params}
\end{adjustwidth}
\end{table}

\nolinenumbers

%
%
%

\bibliography{library}


\end{document}